\documentclass[aps,prl,reprint,superscriptaddress,floatfix,longbibliography]{revtex4-2} 

\usepackage{xspace}
\newcommand{\pentatrap}{\textsc{Pentatrap}\xspace}
\usepackage{color}

\usepackage[normalem]{ulem}

\usepackage{tikz}
\usepackage{pgfplots}
\usepgfplotslibrary{groupplots}
\usetikzlibrary{arrows.meta}%
\usepgfplotslibrary{dateplot}
\usetikzlibrary{shapes,arrows, patterns}%
\usepgfplotslibrary{colormaps}%
\pgfplotsset{compat=newest}
\usetikzlibrary{pgfplots.groupplots,
               pgfplots.fillbetween}
\usetikzlibrary{decorations.pathreplacing}
\usetikzlibrary{decorations.pathreplacing}
\usepgfplotslibrary{dateplot}

\usepackage{hyperref, url}

\def\equationautorefname~#1\null{Eq.~(#1)\null}

\begin{document}


\title{Observation of a low-lying metastable electronic state in highly charged lead by Penning-trap mass spectrometry}

\author{Kathrin Kromer}
\email[]{Corresponding author.\\ kromer@mpi-hd.mpg.de}
\affiliation{Max-Planck-Institut für Kernphysik, 69117 Heidelberg, Germany}
\author{Chunhai Lyu}
\affiliation{Max-Planck-Institut für Kernphysik, 69117 Heidelberg, Germany}
\author{Menno Door}
\affiliation{Max-Planck-Institut für Kernphysik, 69117 Heidelberg, Germany}
\author{Pavel Filianin}
\affiliation{Max-Planck-Institut für Kernphysik, 69117 Heidelberg, Germany}
\author{Zoltán Harman}
\affiliation{Max-Planck-Institut für Kernphysik, 69117 Heidelberg, Germany}

\author{Jost Herkenhoff}
\affiliation{Max-Planck-Institut für Kernphysik, 69117 Heidelberg, Germany}
\author{Paul Indelicato}
\affiliation{Laboratoire Kastler Brossel, Sorbonne Université, CNRS, ENS-PSL Research University, Collège de France, Paris, France}
\author{Christoph H. Keitel}
\affiliation{Max-Planck-Institut für Kernphysik, 69117 Heidelberg, Germany}
\author{Daniel Lange}
\affiliation{Max-Planck-Institut für Kernphysik, 69117 Heidelberg, Germany}
\author{Yuri N. Novikov}
\affiliation{Department of Physics, St Petersburg State University, St Petersburg 198504, Russia}
\affiliation{NRC “Kurchatov Institute”-Petersburg Nuclear Physics Institute, Gatchina 188300, Russia}
\author{Christoph Schweiger}
\affiliation{Max-Planck-Institut für Kernphysik, 69117 Heidelberg, Germany}
\author{Sergey Eliseev}
\affiliation{Max-Planck-Institut für Kernphysik, 69117 Heidelberg, Germany}
\author{Klaus Blaum}
\affiliation{Max-Planck-Institut für Kernphysik, 69117 Heidelberg, Germany}

\date{\today}

\begin{abstract}
Highly charged ions (HCIs) offer many opportunities for next-generation clock research due to the vast landscape of available electronic transitions in different charge states. The development of XUV frequency combs has enabled the search for clock transitions based on shorter wavelengths in HCIs. However, without initial knowledge of the energy of the clock states, these narrow transitions are difficult to be probed by lasers. In this Letter, we provide experimental observation and theoretical calculation of a long-lived electronic state in Nb-like Pb$^{41+}$ which could be used as a clock state. With the mass spectrometer \pentatrap, the excitation energy of this metastable state is directly determined as a mass difference at an energy of $31.2(8)$\,eV, corresponding to one of the most precise relative mass determinations to date with a fractional uncertainty of $4\times10^{-12}$. This experimental result agrees within $1\,\sigma$ with two partially different \textit{ab initio} multi-configuration Dirac-Hartree-Fock calculations of $31.68(13)$\,eV and $31.76(35)$\,eV, respectively. With a calculated lifetime of 26.5(5.3) days, the transition from this metastable state to the ground state bears a quality factor of $1.1\times10^{23}$ and allows for the construction of a HCI clock with a fractional frequency instability of $<10^{-19}/\sqrt{\tau}$. 
\end{abstract}
\maketitle  


The invention of the frequency comb~\cite{hallNobelLectureDefining2006,hanschNobelLecturePassion2006} opened up the possibility to use optical transitions as frequency standards, called optical atomic clocks. These clocks, using single ions in Paul traps~\cite{huntemannSingleIonAtomicClock2016,brewer27AlQuantumLogic2019} or arrays of atoms in optical lattices~\cite{zhengDifferentialClockComparisons2022, bothwellResolvingGravitationalRedshift2022}, have reached incredible fractional frequency instabilities of below $2 \times 10^{-16}/\sqrt{\tau}$ for individual optical clocks~\cite{clementsLifetimeLimitedInterrogationTwo2020}, $\tau$ being the averaging time, and $5 \times 10^{-18}/\sqrt{\tau}$ for spatially separated atomic ensembles~\cite{bothwellResolvingGravitationalRedshift2022}. This extremely high precision not only provides a standard for frequency measurements, it also enables the search for physics beyond the Standard Model, such as temporal or spatial variation of fundamental constants~\cite{prestageAtomicClocksVariations1995, godunFrequencyRatioTwo2014} or violations of Einstein’s equivalence principle through tests of local
position invariance~\cite{langeImprovedLimitsViolations2021}. It is therefore of great importance to push this precision frontier even further to be able to limit the possible size of these effects. One way of achieving this would be to go to strongly forbidden atomic transitions in the extreme ultraviolet (XUV) region. 

With a new generation of frequency combs now spanning more than the optical range and reaching up to the XUV~\cite{gohleFrequencyCombExtreme2005, jonesPhaseCoherentFrequencyCombs2005,pupezaExtremeultravioletFrequencyCombs2021,carstensHighharmonicGeneration2502016, pupezaCompactHighrepetitionrateSource2013,sauleHighfluxUltrafastExtremeultraviolet2019}, research into matching transitions has intensified. Transitions in highly charged ions (HCIs) have become of interest for a new generation of clocks~\cite{crespolopez-urrutiaFrequencyMetrologyUsing2016a, lyuInterrogatingTemporalCoherence2020,lyuinprep}. Though these clocks usually do not hold appropriate transitions for laser cooling and state readout, they can be interrogated via quantum logic spectroscopy~\cite{mickeCoherentLaserSpectroscopy2020}, which has been demonstrated in highly charged Ar$^{13+}$ with a fractional instability of $3\times 10^{-14}/\sqrt{\tau}$ for an optical clock transition~\cite{kingOpticalAtomicClock2022a}. 
 
HCI transitions are widely shielded from field-induced frequency shifts. The remaining electrons of HCIs are bound orders of magnitude stronger than the corresponding electrons in a neutral atom or singly charged ion, making the influence of external fields as well as black-body radiation minimal~\cite{kozlovHighlyChargedIons2018, yudinMagneticDipoleTransitionsHighly2014}. However, the frequency of a coherent laser usually has a limited tuning range, and a metastable state has a narrow transition. Without initial knowledge of the energy of the clock state, it is difficult to design a direct laser-spectroscopy experiment of the clock transition. High-precision mass spectrometry thus provides an alternative approach to infer the energy of a long-lived clock state and, at the same time, to test state-of-the-art theoretical calculations. 

In this Letter, with the Penning-trap mass-spectrometer \pentatrap and the multiconfiguration Dirac-Hartree-Fock (MCDHF) theory, we present the experimental observation and the theoretical calculation of a low-lying metastable electronic state in Nb-like $^{208}$Pb$^{41+}$. The excitation energy is determined to be around 31\,eV with sub-eV uncertainty. Unlike previous mass-spectroscopy measurements of a metastable state around 200\,eV in Re$^{29+}$~\cite{schusslerDetectionMetastableElectronic2020}, this metastable state can be effectively probed via available XUV frequency combs, rendering it more feasible to construct an XUV clock. With a calculated lifetime of $26.5(5.3)$\,days, such a clock bears a quality factor of $1.1\times10^{23}$. Assuming a clock interrogation linewidth of 1\,mHz available for optical lasers~\cite{mateiTextTextEnsuremath2017}, such an XUV clock could achieve a fractional frequency instability of around $4\times10^{-20}/\sqrt{\tau}$. Similar clock states can also be scaled to higher and lower transition energies by employing different elements of the Nb-like isoelectronic sequence~\cite{lyuinprep}.  

To produce highly charged ions, a Heidelberg compact electron beam ion trap (EBIT)~\cite{mickeHeidelbergCompactElectron2018} equipped with an in-trap laser desorption setup is used~\cite{schweigerProductionHighlyCharged2019}. Inside this table-top-sized EBIT, a few-keV electron beam is produced and then guided and compressed by permanent magnets to ionize atoms/ions to high charge states by electron impact ionization. A target made of $^{208}$Pb, positioned next to the trap center, is used as a source of neutral material. By aiming a pulsed Nd:YAG-laser with a pulse energy/duration of about $0.5$\,mJ/8\,ns at the target, a small amount of target material is evaporated into the trap region. In a process called "charge breeding", the ions remain trapped inside the central drifttube, which is set to a depth of 20\,V compared to the neighbouring drifttubes of the EBIT and reach higher and higher charge states by electron impact ionization until they arrive at an equilibrium charge distribution. During electron-ion interaction processes inside the EBIT's ion plasma, such as electron impact excitation or radiative/dielectric recombination, some of the ions' electrons become highly excited~\cite{masseyPropertiesNeutralIonized1942}. After the short-lived excited states decay cascadingly, a fraction of the HCIs will remain in long-lived states for an extended period of time. This fortuitous population of the metastable state allows us to measure the excitation energy of the metastable state without having to actively drive the transition by using mass measurements. This method does not require prior knowledge of the state energies, which makes these measurements independent of theory or other experiments. 

The HCIs are extracted from the EBIT and slowed down from 4\,keV$/q$ to a few eV$/q$ to be able to trap them in a cryogenic Penning trap. The ion transport at the experiment \pentatrap works via an electrostatic beamline in combination with a Bradbury Nielson Gate~\cite{bradburyAbsoluteValuesElectron1936} for charge state selection. The ions are then decelerated by a set of two pulsed drift tubes, one of which is situated in the room temperature region (deceleration down to $\approx 200$\,eV$/q$) and a second one in the cryogenic part of the beamline (down to a few eV$/q$) and ultimatively captured in the first Penning trap. For an overview of the beamline setup, see~\cite{kromerHighprecisionMassMeasurement2022}.

\begin{figure*}
    \centering
    \includegraphics[width=165mm]{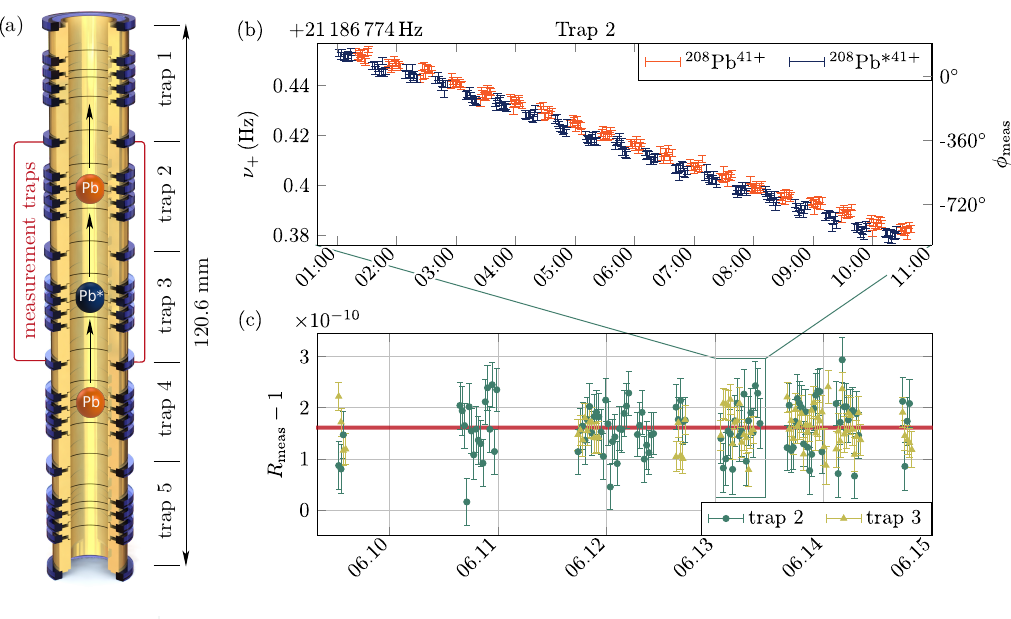} 
    \caption{a) Schematic drawing of the Penning-trap tower with three ions in configuration 1. The ion transport to configuration 2 is implied by arrows. The phase measurements are carried out in traps 2 and 3. Traps 1 and 4 are utilized as storage traps and trap 5 is currently not in use. b) Example of $\nu_+$ data from a measurement in trap 2 with $^{208}$Pb$^{41+}$ ions, one being in the ground state (orange) and one in the metastable state (blue). On the left $y$-axis, the resulting frequency is shown while on the right $y$-axis, the unwrapped, measured phase $\phi_{\text{meas}}$ is given. When combining the averaged modified cyclotron frequency with the axial and magnetron frequencies, see \autoref{eq:invariance}, one can calculate the cyclotron frequencies and their ratios $R$ plotted below in subfigure c). This plot shows the ratios of all relevant measurement runs in both traps. The red line gives the average and its width corresponds to the combined statistical and systematic error band. }
    \label{fig:results_traptower}
\end{figure*}

To determine mass ratios of stable or long-lived highly charged ions, five identical, aligned, cylindrical Penning traps are used in \pentatrap's trap tower, see \autoref{fig:results_traptower}a)~\cite{reppPENTATRAPNovelCryogenic2012,rouxTrapDesignPENTATRAP2012}. The trap tower opens the possibility to conduct simultaneous measurements on two ions stored in traps 2 and 3 while having two more traps for ion storage. The fifth trap is currently not in use. The measurement principle is based on a measurement of the frequencies of the ions' three Penning-trap eigenmotions: the modified cyclotron frequency $\nu_+$, the axial frequency $\nu_z$, and the magnetron frequency $\nu_-$ and calculating the free cyclotron frequency $\nu_c = \frac{q}{2\pi m}B$ by applying the invariance theorem~\cite{brownPrecisionSpectroscopyCharged1982}
\begin{equation}
    \nu_c = \sqrt{\nu_+^2 +\nu_z^2 + \nu_-^2} \text{ .}
     \label{eq:invariance}
\end{equation} 
The free cyclotron frequency $\nu_c$ is inversely proportional to the mass $m$ of the ion and proportional to the exactly known charge $q$ of the ion and the magnetic field $B$. In order to achieve a precise determination of the mass of the ion in the excited electronic state $m_{\text{exc}}$, the cyclotron frequency of an ion in the ground state $\nu_{c\text{,g}}$ is measured alternately to that of an ion in the excited state $\nu_{c\text{,exc}}$ in each of the measurement traps. When determining the ratio $R = \frac{\nu_{c\text{,g}}}{ \nu_{c\text{,exc}}}$ for trap 2 or trap 3 of the two cyclotron frequencies, the magnetic field, being the least precisely known quantity, cancels out to first order and the identical charge of both ions drops out of the ratio. The mass difference $\Delta m$ between the ion in the ground state $m_{\text{g}}$ and the excited state $m_{\text{exc}}$ can then be determined with
\begin{equation}
    \Delta m = m_{\text{exc}}-m_{\text{g}} 
    = m_{\text{g}} (R-1) \text{ .}
    \label{eq:Ratio_determination}
\end{equation}

\begin{figure}[b]
    \centering
    \includegraphics[width= 86mm]{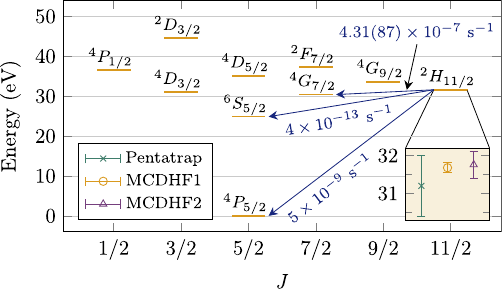}
    \caption{Lowest energy levels of the Nb-like Pb$^{41+}$ ion. The metastable [Kr]$4d^5$ $^2H_{11/2}$ state lies at around 31\,eV above the [Kr]$4d^5$ $^4P_{5/2}$ state. It decays mainly via an $E2$ transition to the short-lived state [Kr]$4d^5$ $^4G_{7/2}$. The metastable state has a lifetime of 26.5(5.3)\,days. }
    \label{fig:levelplot}
\end{figure}

The individual ions in each measurement trap, see \autoref{fig:results_traptower}a), are detected non-destructively using the Fourier-transform ion-cyclotron-resonance technique (FT-ICR)~\cite{fengTankCircuitModel1996}. This well-established method uses the ion's image current, which is converted into a measurable voltage drop across a resonant tank circuit at cryogenic temperatures. When the cryogenic tank circuit is connected to an axially offset electrode, one can reduce the amplitude of the axial motion by thermalizing it, and measure its frequency as a 'dip' in the resonance curve~\cite{fengTankCircuitModel1996}. By coupling the modified cyclotron or the magnetron motion to the axial motion, the coupled motion is cooled and its frequency can be measured using the double-dip technique~\cite{cornellModeCouplingPenning1990b}. This technique is applied to determine the magnetron frequency and for estimating frequencies and coupling pulse times for the phase-sensitive Pulse and Phase (PnP) method used in the main measurements~\cite{cornellSingleionCyclotronResonance1989}. A PnP cycle works as follows: in the beginning, the cyclotron motion is excited using a dipolar radio-frequency (RF) pulse, to set a specific starting phase; then the ion's cyclotron oscillation is left to evolve freely for a well-known time $t_{\text{acc}}$, called phase accumulation time. The energy and phase information of the cyclotron mode is then coupled to the axial mode using a $\pi$ pulse at the sideband frequency $\nu_{\text{rf}} = \nu_+ - \nu_z$. Subsequently, the cyclotron phase information can be read out using the axial image-current detection system. With this method, one can distinguish between the ground state and low-lying metastable state in $^{208}$Pb$^{41+}$, see \autoref{fig:levelplot}. At a phase accumulation time of $t_{\text{acc}} = 40$\,s the phase difference between metastable and ground state of about $49^{\circ}$ at $\nu_+ \approx 21.2$\,MHz exceeds the phase stability of about $12^{\circ}$, see \autoref{fig:results_traptower}b).

Once there are three ions loaded and identified in the alternating configuration shown in \autoref{fig:results_traptower}a), the PnP measurement loop is started. In addition to the prior mentioned phase measurement, this also includes an axial frequency measurement during the phase accumulation time of the modified cyclotron frequency and a single magnetron frequency measurement in the preparation phase of the measurement. During the continuous PnP measurement loop, the ions are moved up or down every 10 measurement points to alternate between configurations 1 and 2 effectively swapping the ions in each measurement trap from the ion in the ground state to the ion in the excited state or vice verca, see \autoref{fig:results_traptower}. For further information on the measurement procedure, see \cite{rischkaMassDifferenceMeasurementsHeavy2020, schusslerDetectionMetastableElectronic2020, filianinDirectValueDetermination2021,kromerHighprecisionMassMeasurement2022}. 

In the analysis, all three measured eigenfrequencies are combined, following \autoref{eq:invariance}, and the cyclotron frequency ratio is formed by interpolation, whereby one of the ion's block of 10 cyclotron frequency measurements is averaged and interpolated to the time of the other ion's averaged cyclotron frequency measurement. In \autoref{fig:results_traptower}b) it is clearly visible that the modified cyclotron frequency $\nu_+$ drifts downward by $\delta B/B =-2.3 \times 10^{-10}$\,/h due to a loss of magnetic field of the superconducting magnet. With the interpolation, this magnetic field drift is cancelled out to first order.

The resulting ratios of both traps can be found in \autoref{fig:results_traptower}c). The average ratio of all measurement runs was determined to be $R_{stat}-1 = 1.609(32) \times 10^{-10}$. Due to the almost identical charge-to-mass ratio of the ion in ground and metastable state, most known systematic effects, such as the image charge shift or the relativistic shift, cancel out to a large extent when forming the frequency ratio. Only one systematic uncertainty of relevant size remains, namely the dip-lineshape uncertainty~\cite{rauPenningTrapMass2020}. This systematic originates from the uncertainty of the axial dip frequency relative to the fitted resonator frequency. In trap 3, the $Q$-factor of the resonator is substantially larger and thus the dip width is substantially broader than in trap 2, which increases the uncertainty of the resonator frequency because the center of the resonance spectrum is "covered" by the broad dip. As expected, the fit of the dip in trap 3 yields a larger systematic error associated with the mismatch of the dip and resonator frequencies connected to the distortion of the dip shape due to the frequency pulling effect. These two effects in trap 2 and trap 3 result in an error of $<1\times 10^{-12}$ and $3.3\times 10^{-12}$, respectively. Including this systematic effect, the final experimental mass ratio is determined to be $R-1 = 1.61(4) \times 10^{-10}$. The mass or energy difference of the two states can then be calculated using \autoref{eq:Ratio_determination} to be $31.2(8)$\,eV. This small energy difference was measured as a relative mass measurement against the total mass of the Pb ion of $\approx$194\,GeV/c$^2$, reaching a relative precision of $4\times 10^{-12}$.

The level structure and lifetimes of the lowly-lying states in Pb$^{41+}$, see \autoref{fig:levelplot}, are calculated with the \textit{ab initio} multiconfiguration Dirac--Hartree--Fock (MCDHF) and relativistic configuration interaction (RCI) methods in two partially different implementations, one of them being the GRASP2018 code~\cite{Grant1970,Desclaux1971,GRASP2018} (referred to henceforth as MCDHF1) and the second one based on the MCDFGME code~\cite{MCDFGME} (referred to as MCDHF2). With a ground state of [Kr]$4d^5$ $^4P_{5/2}$, the metastable state is determined to be $^2H_{11/2}$. 

Within the calculations of MCDHF1, each many-electron atomic state function (ASF) is expanded as a linear combination of configuration state functions (CSFs) with common total angular momentum ($J$), magnetic ($M$), and parity ($P$) quantum \mbox{numbers:} $|\Gamma P J M\rangle = \sum_{k} c_k |\gamma_k P J M\rangle$. The CSFs $|\gamma_k P J M\rangle$ are $jj$-coupled Slater determinants of one-electron orbitals and $\gamma_k$ summarizes all the remaining information needed to fully define the CSF, i.e., the orbital occupation and coupling of single-electron angular momenta. $\Gamma$ collectively denotes \mbox{all} the $\gamma_k$ included in the representation of the ASF. 

In the calculation of the excitation energy of the $J=11/2$ metastable state, the CSF basis set is generated via single and double (SD) excitation of electrons from the $4s^24p^64d^5$ reference configurations to high-lying virtual orbitals. After obtaining the radial wave functions from the self-consistent MCDHF calculations under the Dirac--Coulomb Hamiltonian, the RCI method is applied to derive the mixing coefficients $c_k$ and excitation energies, as well as the approximate corrections arising from the mass shift, quantum electrodynamic (QED) terms and Breit interactions. 

To monitor the convergence of the result, we systematically added and optimized virtual orbitals layer by layer up to $n=8$ with all orbital quantum numbers ranging from 0 to $n-1$ being included ($n$ is the principal quantum number). Through extrapolation to $n=\infty$, one obtains an energy of $31.672$\,eV within the calculation of MCDHF1 with 8.92 million CSFs, which includes a $837$\,meV contribution from the Breit interactions, $47.2$\,meV from the self-energy corrections, $0.07$\,meV from the vacuum polarization, 1.72\,meV from the field shift, and $0.24$\,meV from the mass shift. The RCI calculations based on different radial wave functions lead to a correction of 6\,meV. By varying the fine-structure constant in the calculation, this transition is found to have an $\alpha$-variation sensitivity of $K=(\Delta~E/E)/(\Delta\alpha/ \alpha)=-1.74$. Considering the ten times greater transition energy, the absolute frequency change due to $\alpha$ variations is comparable to other HCI candidates.

Furthermore, to account for the core--core correlations, the CSFs generated via SD excitations from the $1s$ orbital up to $5g$ orbital are added to the RCI calculations. This gives rise to a correction of $-20$\,meV to the previous value. Then, through further inclusion of SD excitations from multi--reference $\{4s4p^64d^6,~4p^64d^7,~4s^24p^44d^7\}$ configurations up to the $6h$ orbital, the contribution from dominant triple and quadruple electron exchanges is considered in the RCI procedure with 13.7 million CSFs. This gives a correction of $14$\,meV from high-order electron correlations. The final excitation energy is determined to be $31.68(13)$\,eV. The uncertainty is conservatively given as the absolute summation of the corrections from the core--core and high-order correlations, the radial wave function dependence, plus 10\% uncertainty from the Breit and QED terms. 
 
The approach of MCDHF2 deviates from the above described calculations in the following way: In contrast to the perturbative approach of MCDHF1, the calculation within MCDHF2 includes the Breit and Retardation term of the electron-electron interaction to all order by adding it directly to the many-body Hamiltonian. A smaller set of configurations is used compared to MCDHF1, but including a full relaxation of all orbitals. Single, double, and triple excitations were limited to excitations from the $n=4$ shell: $4s$, $4p$, $4d$ to $4d$, $4f$, $5s$, $5p$, $5d$,$ 6s$, $6p$ with approximately 29\,000 configurations for the $^2H_{11/2}$ state and 24\,000 configurations for the ground state. For the self-energy screening, two different methods are used~\cite{indelicatoMulticonfigurationalDiracFockStudies1987,indelicatoMulticonfigurationDiracFockCalculations1990,shabaevModelOperatorApproach2013} and more high-order QED corrections are included, however, the contribution of this is minimal for the case of transition energies in Pb$^{41+}$. The MCDHF2 method reaches a final value for the excitation energy of $31.76(35)$\,eV, in excellent agreement with the MCDHF1 result.

While the metastable state decays to the ground state via a magnetic octupole ($M3$) transition under a rate of $A_1 = 4.95(5)\times10^{-9}$\,s$^{-1}$, its lifetime is mainly determined by the electric dipole ($E2$) decay channel to the $J=7/2$ state, see \autoref{fig:levelplot}. The decay rates were calculated with the MCDHF1 method. Under the above-mentioned multi--reference scheme, we expanded the CSF basis set to $n=8$ and obtained a rate of $5.18\times10^{-7}$ and $4.31\times10^{-7}$\,s$^{-1}$ in the Coulomb and Babushkin gauge, respectively. Since both rates show decreasing trends with $n$ and the rate in the Babushkin gauge is already close to convergent, the value $A_2 = 4.31(87)\times10^{-7}$\,s$^{-1}$ is used to represent the decay rate of this channel. The uncertainty is given as the difference between the two rates in the different gauges. As a result, the lifetime of the metastable state is determined to be 26.5(5.3)\,days.

The resonant photon-excitation cross section~\cite{foot2004atomic} from the ground state to this metastable state is calculated to be $8.3\times10^{-14}$\,cm$^{2}$ per photon. Current XUV frequency combs at such photon energy would have a repetition rate of 100\,MHz and a pulse duration of 24\,fs~\cite{nautaXUVFrequencyComb2021}. With proper phase-matching schemes, XUV combs with powers at the mW level are achievable~\cite{porat2018phase}. Assuming an average power of 5\,mW per harmonic or 25.5\,nW per tooth and a focal size of 10\,$\mu$m$^2$, they bear a photon flux of $5.0\times10^{16}$\,ph./s/cm$^2$ per tooth. Assuming a comb coherence time of 1\,s~\cite{benko2014extreme}, i.e., a tooth width of 160\,mHz, the resonant photon flux would be $2.2\times10^{10}$\,ph./s/cm$^2$. Thus, one would obtain $1.8\times10^{-3}$ excitations per second (or an effective Rabi frequency of $\Omega/2\pi=6.8$~mHz) at resonance~\cite{lyuInterrogatingTemporalCoherence2020}, rendering future XUV resonant spectroscopy of this clock state promising. Nevertheless, before direct excitation of the metastable state, further spectroscopy of the nearby fast transitions such as $^4P_{5/2} \rightarrow $ $^4G_{7/2}$, $^4G_{7/2} \rightarrow $ $^4G_{9/2}$, and $^4G_{9/2} \rightarrow $ $^2H_{11/2}$ is necessary to improve the accuracy of the clock transition energy to around 21~kHz. For the quantum logic scheme, however, one needs to be able to drive a sideband coupling in the HCI, whose Rabi frequency is typically smaller by a factor of $\eta=kz_0$. Here, $\eta$ is the Lamb--Dicke parameter, $k$ the wavenumber of the laser and $z_0$ the spatial spread of the ground state motional wave function along the laser propagation direction~\cite{winelandExperimentalIssuesCoherent1998}. Therefore, in order to realize the full ability of quantum logic, one needs to either increase the laser power or the coherence time of the XUV comb. 

We have shown the experimental determination and theoretical calculation of the excitation energy of a low-lying metastable state in Pb$^{41+}$. The metastable [Kr]$4d^5$ $^2H_{11/2}$ state and its most probable decay channels including the transition rates, calculated with the MCDHF1 method, are shown in \autoref{fig:levelplot}, and the region of interest around the metastable state is enlarged to show the different values of experiment (\pentatrap) and theory (MCDHF1 and 2). As can be seen, the calculation of MCDHF1 and MCDHF2 agree with the experimental data within $1\,\sigma$. This comparison verifies the theoretical multi-electron correlation studies and excitation energy estimations described in this Letter. It also shows the ability of Penning-trap mass measurements to help in the search for transitions usable in future HCI clocks and to determine their energy at a sub-eV precision.

As described, the energy of the metastable state falls into the range of current XUV combs, allowing resonant spectroscopy of this metastable state using a high flux, high repetition rate XUV comb~\cite{sauleHighfluxUltrafastExtremeultraviolet2019}. The orders-of-magnitude higher transition energy of XUV in comparison to optical clocks could enable the construction of an ultrastable clock with a fractional frequency instability around $4\times10^{-20}/\sqrt{\tau}$.

\begin{acknowledgments}
This work comprises parts of the Ph.D. thesis work of K.K. to be submitted to Heidelberg University, Germany.
This work is part of and funded by the Max-Planck-Gesellschaft and the DFG (German Research Foundation) – Project-ID 273811115 – SFB 1225 ISOQUANT. The project received funding from the European Research Council (ERC) under the European Union’s Horizon 2020 research and innovation programme under grant agreement number 832848 - FunI. P. I., Y. N., and K. B. are members of the Allianz Program of the Helmholtz Association, contract n° EMMI HA-216 “Extremes of Density and Temperature: Cosmic Matter in the Laboratory.” Furthermore, we acknowledge funding and support by the International Max-Planck Research School for Precision Tests of Fundamental Symmetries and the Max Planck, RIKEN, PTB Center for Time, Constants and Fundamental Symmetries. The authors thank J. R. Crespo López-Urrutia and his team for insightful discussions and support.
\end{acknowledgments}

\bibliography{metastable}

\end{document}